\documentstyle[aps,epsf]{revtex}

\begin{document}

\draft
\preprint{}  

\title{The origin of scale-scale correlations of the
density perturbations during inflation}
\author{Li-Zhi Fang$^1$, Wolung Lee$^1$ and Jes\'us Pando$^2$}
\address{$^1$ Department of Physics, University of Arizona,
         Tucson, AZ 85721, USA}
\address{$^2$ UMR 7550 CNRS, Observatoire de Strasbourg,
        11 Rue de l'Universit\'e, 67000 Strasbourg, France}
\date{\today}

\maketitle

\begin{abstract}

We show that scale-scale correlations are a generic feature of slow-roll
inflation theories. These correlations result from the long-time tails 
characteristic of the time dependent correlations because the long wavelength 
density perturbation modes are diffusion-like. A relationship 
between the scale-scale correlations and time-correlations is established
providing a way to reveal the time correlations of the perturbations
during inflation. This mechanism provides for a testable prediction that 
the scale-scale correlations at two different spatial points 
will vanish.

\end{abstract}

The inflation paradigm for the evolution of the universe not only solves
classical cosmological dilemmas such as the horizon and flatness problems, 
but also provides the seeds for structure formation in the form of quantum
or thermal fluctuations\cite{KT}.  These fluctuations grow by gravitational 
instability to form the structure we see today.  The nature of the 
fluctuations leads to the belief that the primordial density field is
likely Gaussian. However, recent analysis of Ly$\alpha$ forest lines 
in quasar absorption spectra reveal that the Ly$\alpha$ clouds at redshifts 
between 2 and 3 are significantly scale-scale correlated on scales of 40 to
80 $h^{-1}$ Mpc\cite{PLGF}. Since the gravitational clustering of the
universe remains in the linear regime on scales larger than about
30 $h^{-1}$ Mpc at redshifts higher than 2, this result implies that the 
primordial density fluctuations of the universe are probably scale-scale 
correlated. More direct evidence for the scale-scale correlation in the 
primordial density field has recently been found in the Cosmic Background 
Explorer (COBE) Differential Microwave Radiometer (DMR) 4-year all 
sky maps\cite{PVF}.  In particular, the observed cosmic
temperature fluctuations on the North Galactic Pole are found, with
$>99\%$ confidence, to be scale-scale correlated on angular scales of
10 to 20 degrees.  These scales are much larger than the Hubble size at
the decoupling era and cannot be attributed to any known foregrounds or 
correlated noise maps. Moreover, the correlation is on the order of
$\simeq 10^{-5}$ K, i.e., the same order as the temperature fluctuations,
and therefore it is unlikely to be a higher order effect.

Although confirmation of scale-scale correlations in the primordial
density field must await further observations, it is appropriate now
to study the possible origin of these correlations. At the first glance, these
correlations might appear to be incompatible with inflation.  However, in 
this {\em Letter}, we show that scale-scale correlations are not only 
possible, but a generic feature of the inflationary scenario. 

It is known that there are models 
which lead to a density field with a Poisson or Gaussian distribution in the 
one-point distribution functions, but that are highly scale-scale 
correlated\cite{g95a}. The physical reason for the co-existence of Gaussian 
one-point functions and scale-scale correlations can easily be seen in phase 
space, i.e., position-wavevector space. Any density field $\rho({\bf r})$ can 
be decomposed by an orthogonal and complete basis $\Psi_{{\bf k},{\bf x}}$,
where the indexes ${\bf k}$ and ${\bf x}$ denote, respectively, the
wavevector and position of a volume element $d^3 x d^3 k \simeq 1$ in
phase space. One way to realize this decomposition is by the discrete wavelet 
transform (DWT)\cite{FT}. With this type of decomposition, the behavior of 
$\rho({\bf r})$ in phase space is described by the projection of 
$\rho({\bf r})$ on $\Psi_{{\bf k},{\bf x}}$, which we denote by
$\tilde{\epsilon}_{{\bf k},{\bf x}}$. The density perturbations of component
$\tilde{\epsilon}_{{\bf k},{\bf x}}$ are localized at ${\bf k},{\bf x}$ in 
phase space, and thus one can have distributions for which 
$\tilde{\epsilon}_{{\bf k},{\bf x}}$ is Gaussian in its one-point distribution
with respect to ${\bf x}$ while scale-scale correlated in terms of 
${\bf k}$. For instance, consider a distribution of 
$\tilde{\epsilon}_{{\bf k},{\bf x}}$ which is Gaussian in 
${\bf x}$ at a given scale $|{\bf k}|=k_1$.  Then
\begin{equation}
\langle \tilde{\epsilon}_{{\bf k},{\bf x}}
\tilde{\epsilon}_{{\bf k},{\bf x'}} \rangle = 
P_{dwt}(k_1)\delta_{{\bf x},{\bf x'}},
\end{equation}
where $P_{dwt}(k_1)$ is the variance or the power spectrum with respect
to the DWT decomposition \cite{PF}. All higher order correlations
of $\tilde{\epsilon}_{{\bf k},{\bf x}}$ are zero at this scale. If 
perturbation $\tilde{\epsilon}_{{\bf k},{\bf x}}$ at scale 
$|{\bf k}|=k_2$ is related to that at scale $|{\bf k}|=k_1$ as
\begin{equation}
\tilde{\epsilon}_{|{\bf k}|=k_2,{\bf x}}= \alpha
\tilde{\epsilon}_{|{\bf k}|=k_1,{\bf x}}, 
\end{equation}
where $\alpha$ is a constant, the distribution of
$\tilde{\epsilon}_{{|\bf k}|=k_2,{\bf x}}$ is also Gaussian in ${\bf x}$ 
with variance $P_{dwt}(k_2)=\alpha^2P_{dwt}(k_1)$, but is
strongly scale-scale correlated. That is, at a given position ${\bf x}$, 
the scale $k_2$ perturbation is always proportional to the scale $k_1$
perturbation. Scale-scale correlations depend only on the statistical 
behavior of the fluctuation distribution along the ${\bf k}$ direction 
but localized in the ${\bf x}$ direction of the phase space. Obviously
such scale-scale correlations can only be effectively detected by a
scale-space decomposition.

In the inflationary scenario, the statistical behavior of the primeval density 
perturbations in the ${\bf k}$ direction are determined by the time-dependent
correlations of the fluctuations during inflation. This is because the density
perturbations on comoving wavenumber $k$ originate from fluctuations 
crossing over the Hubble radius at time $t_c$, and 
\begin{equation}
k \propto e^{Ht_c},
\end{equation}  
where $H$ is the Hubble constant during inflation. Equation (3) is a well
known feature of inflation: the larger the scale of the perturbations,
the earlier the time of the formation\cite{KT}. Equation (3) is actually a
mapping of time $t$ to scale space $k$ of the density perturbations. Thus, 
the time-time correlations during the inflation,
$\langle \phi(t)\phi(0) \rangle$, where $\phi(t)$ denotes fluctuations
of the inflaton $\Phi$, will be mapped into scale-scale correlations.

The primeval perturbations will not be scale-scale correlated if the 
time dependent correlation functions decay faster than Hubble time $H^{-1}$.
In general, the time-time correlation functions of fluctuations in 
equilibrium decay exponentially within the mean-free time $t_f$ 
between collisions, which is not longer than about $1/H$. Such fluctuations
do not lead to scale-scale correlations. However, it is well known both
experimentally and theoretically that there is a generic mechanism for
long-range correlations in {\em nonequilibrium} systems. For these systems,
the temporal correlation functions of perturbations decay only as a
power-law,
\begin{equation}
\langle\phi(t) \phi(0) \rangle \propto (t_f/t)^{d/2},
\end{equation}
where $d$ is the dimension of the system. This long-time tail 
phenomenon is found to be applicable to all systems where the relaxation 
is dominated by long wavelength perturbations via diffusion-like 
dissipation \cite{ehv,KB}. Inflation can be treated as such a system thus 
implying the existence of scale-scale correlations.

We can confirm this by considering a standard slow-roll inflation 
governed by a scalar field $\Phi$ satisfying the equation of motion as
\begin{equation}
\ddot{\Phi} + 3H \dot{\Phi} - e^{-2Ht}\nabla^2\Phi+ V'(\Phi)=0,
\end{equation} 
where $V(\Phi)$ is the potential of $\Phi$. For our purposes, it is
convenient to employ the stochastic inflation formalism \cite{star,LF}, in
which the $\Phi$ field is decomposed into large scale component
$\phi({\bf x},t)$ and small scale modes $q({\bf x}, t)$ as
\begin{equation}
\Phi({\bf x}, t)=\phi({\bf x}, t) + q({\bf x}, t).
\end{equation}
The component $\phi({\bf x}, t)$ is actually a coarse-grained $\Phi$ 
over size $1/k_{c} > 1/H$. It contains the uniform background and
perturbations on scales larger than $1/k_{c}$, and therefore can be
treated classically. (Here and below we use $k_c$ and $k$ to describe 
the wavenumbers of the fluctuations during inflation. It should not be 
confused with the $k$ in Eq. (3), where it is the wavenumber of the 
density perturbations after inflation). The term $q({\bf x}, t)$ contains 
all high frequency quantum or thermal fluctuations. For slow-roll inflation, 
the evolution of $\phi({\bf x}, t)$ modes obeys a Langevin equation
\begin{equation}
\frac {\partial \phi({\bf x}, t)}{\partial t} =  - \frac{1}{3H}
\frac{\delta F[\phi({\bf x})]}{\delta \phi} + 
\eta({\bf x}, t),
\end{equation}
where 
$F[\phi]=\int d^3{\bf x} \left[(e^{-Ht}\nabla\phi)^2/2 +V(\phi) \right]$ is 
the Ginzburg-Landau free energy. The noise term is given by the 
short-wavelength modes
\begin{equation}
\eta({\bf x}, t)=\left (-\frac{\partial}{\partial t}
   +\frac{1}{3H}e^{-2Ht}\nabla^2\right) q({\bf x}, t), 
\end{equation}
and its time correlation function is approximately
\begin{equation}
\langle \eta({\bf x}, t) \eta({\bf x'}, t') \rangle \simeq
D \delta ({\bf x}-{\bf x'})\delta(t-t'),
\end{equation}
where $D=H^3/2\pi$ or $H^2T/2\pi$ ($T$ being temperature of heat bath)
for quantum or thermal fluctuations, respectively. As expected, Eq. (9)
shows that the short wavelength modes rapidly decay and are Gaussian.

However, the primordial density perturbations originate from the long
wavelength modes $\phi$. The time correlation behavior of $\phi$ can be
seen from a mode analysis of Eq. (7). Linearizing Eq. (7), one has
\begin{equation}
\frac {\partial \phi}{\partial t} = 
  -\left [\frac{\phi}{\tau_0} - 
 \frac{e^{-2Ht}}{3H}\nabla^2\phi \right ] + \eta
\end{equation}
with the zero wavenumber relaxation time given by
\begin{equation}
\tau_0^{-1}= V''(0)/3H .
\end{equation}
Taking the Fourier transform of Eq. (10) gives
\begin{equation}
\frac{\partial \hat{\phi}_{\bf k}}{\partial t}= 
- \frac{ \hat{\phi}_{\bf k}}{\tau_{\bf k}} + \eta_{\bf k} ,
\end{equation}
where ${\bf k}$ is a comoving wave-vector. The relaxation time of the mode
${\bf k}$ is 
\begin{equation}
\frac{1}{\tau_{\bf k}} = \frac{1}{\tau_{0}} + \frac{1}{3H}e^{-2Ht}k^2
  =\frac{1}{\tau_{0}} + \frac{k_p^2(t)}{3H},
\end{equation}
where $k_p(t)=k\exp(-Ht)$ is the physical wavenumber of $k$. Equation (13) is a
dispersion relation of mode ${\bf k}$, i.e., $i\omega = \tau^{-1}_{{\bf k}}$. 
If the potential $V$ is flat, $V' \simeq V'' \simeq 0$, we have
$\tau_0^{-1} \simeq 0$, and
$\omega = -ik_p^2(t)/3H$. This is typical of diffusion-damping soft modes
for which the relaxation time
$\tau_{\bf k} \rightarrow \infty$ when $k_p \rightarrow 0$. These modes
lead to long-range correlations in general\cite{KB}.

Since perturbations $\phi({\bf x},t_0)$ only consist of modes with
physical wavelength larger than $1/k_{c}$ at $t_0$, the
Fourier transform of $\phi({\bf x},t_0)$ is limited to these modes as
\begin{equation}
\phi({\bf x}, t_0) = \int_{k_p(t_0) < k_c}
d^{3}{\bf k} \hat{\phi}_{\bf k} e^{-i\omega t_0-i{\bf k}\cdot {\bf x}}.
\end{equation}
We are concerned with perturbations $\phi({\bf x}, t_0)$ localized
within the Hubble radius $1/H$ ($< 1/k_c$) and in this range, the phase factor
$\exp(-i{\bf k}\cdot {\bf x})$ in Eq. (14) is almost a constant. Moreover
$\phi({\bf x}, t_0)$ should be described by an ensemble of solutions of
Eq. (8) with various realizations of $\eta$. Because $\eta$ is white
noise, the amplitude of fluctuations $\hat{\phi}_{\bf k}$ will be
$k$-independent on average, i.e.,
$\langle \hat{\phi}_{\bf k} \hat{\phi}^*_{\bf k} \rangle \simeq
\langle \hat{\phi}^2 \rangle$ is constant, where $\langle ... \rangle $
denotes averaging with respect to the ensemble of $\eta$. Hence, we have
$\langle \phi \phi^*\rangle \equiv
\langle \phi({\bf x}, t_0)\phi^*({\bf x}, t_0) \rangle
\simeq k_c^6 \langle \hat{\phi}^2 \rangle$. The values
$\langle \hat{\phi}^2 \rangle$ for quantum and thermal fluctuations have
been calculated in \cite{star,LF}.

The Fourier transform of $\phi({\bf x}, t)$ is given by
\begin{eqnarray}
\phi({\bf x},t) & = & \int_{k_p(t) < k_{c}} d^{3}{\bf k}
\hat{\phi}_{\bf k} e^{-i\omega t -i{\bf k}\cdot {\bf x} }
\nonumber \\ 
& = &
\int_{k_p(t) < k_{c}} d^{3}{\bf k}
\hat{\phi}_{\bf k} e^{-i\omega t_0 -i{\bf k}\cdot {\bf x} }
\exp \left (-\left
[\frac{1}{\tau_0} + \frac{k_p^2(t)}{3H} \right ](t-t_0)\right ).
\end{eqnarray}
Thus, the time-dependent correlation function is
\begin{equation}
\langle \phi({\bf x}, t)\phi^{*}({\bf x}, t_0) \rangle =  
k_c^3 \langle \hat{\phi}^2 \rangle
\int_{k_p(t) < k_{c}} d^3{\bf k}
\exp \left ( -\left
[\frac{1}{\tau_0}+\frac{k_p^2(t)}{3H}\right ] (t-t_0) \right ).
\end{equation}
In the case of a flat potential $\tau_0^{-1}=0$, we have the long-time 
tail correlation function as
\begin{equation}
\frac{\langle \phi({\bf x}, t)\phi^{*}({\bf x}, t_0) \rangle}
{\langle \phi \phi^* \rangle}
\simeq \left (\frac{t_f}{t-t_0} \right )^{3/2}
\hspace{5mm}{\rm with}\hspace{5mm} (t-t_0) \gg t_f,
\end{equation}
and
\begin{equation}
t_f = \frac{3H}{k^2_c} > \frac {1}{H}.
\end{equation}
Comparing Eq. (17) with Eq. (4), we have as expected $d=3$, and the 
``mean-free time'' is determined by the cut-off $1/k_c$, i.e., the
minimum scale of the fluctuations $\phi$.  

When the potential $V$ is not constant, the time correlation function
$\langle \phi(t)\phi(t_0)\rangle$ [Eq. (17)] contains an exponential
decay factor $\exp[-(t-t_0)/\tau_0] $. However, the slow-roll
dynamics requires that $V'' \ll H^2$. Thus, the factor
$\exp[-(t-t_0)/\tau_0 ]$ can always be ignored as it is equal to about 1
provided $H(t-t_0)$ is not very large. Therefore, the long temporal
correlations do not depend on the details of inflation. In fact, the
slow-roll condition in inflation plays the same role as the critical
slowing down in phase transitions. Under these conditions, the relaxation
is dominated by soft modes and long range temporal correlations 
occur\cite{KB}. Because the background space is uniform, these soft modes 
do not induce long range spatial correlations and therefore, the 
perturbations in different space ${\bf x}$ are uncorrelated. In other words, 
the perturbation field has a Gaussian one-point distribution with respect to
${\bf x}$.

For inflation with thermal dissipation, Eq. (17) still holds \cite{BF}.
In this case $1/k_c$ can be as small as the Hubble size $1/H$\cite{LF}, 
and therefore
\begin{equation}
t_f  \simeq  \frac{1}{H},
\end{equation}
i.e. the mean-free time is given by the Hubble time. 

It is important to stress that although the long temporal correlations
are calculated here using the stochastic formalism, this result is far more 
general than any specific stochastic inflation ansatz. The Langevin equation 
(10) for the long wavelength fluctuations of the scalar field can be derived 
from first principles by considering the quantum evolution of the reduced 
density matrix. Since it is necessary that quantum coherence be lost in order 
to treat the initial density perturbations classically with general relativity,
a quantum-to-classical decoherence is necessary. The decoherence via a 
coarse-grained quantum temporal evolution produces classical correlations in 
phase space like Eq. (17) through the interaction between long wavelength 
modes and the short wavelength ``thermal'' noise of the Hawking temperature 
$H$\cite{nam}. Thus long temporal correlations are a generic feature of 
inflationary models with a slow-roll regime.

Since the evolution of perturbations from horizon-crossing to the time of 
decoupling is linear, we have
\begin{equation}
\tilde{\epsilon}_{k, {\bf x}} \propto \phi({\bf x}, t_c),
\end{equation} 
where $\tilde{\epsilon}_{k,{\bf x}}$ is the wavelet coefficient of density or 
temperature fluctuations on scale $k$ at position ${\bf x}$, and $k$ is 
related to $t_c$ via Eq. (3). Thus, the scale-scale correlations of the 
cosmic temperature fluctuations should be
\begin{equation}
\langle \tilde{\epsilon}_{k_1,x_1}\tilde{\epsilon}_{k_2,x_2}\rangle
\propto
\langle\phi(t_{c1})\phi^{*}(t_{c2})\rangle.
\end{equation}

The scale-scale correlations of the cosmic temperature fluctuations 
have been detected using 
\cite{PLGF}
\begin{equation}
C(k_1,k_2,x) \equiv
\frac{\langle \tilde{\epsilon}^2_{k_1,x}\tilde{\epsilon}^2_{k_2,x}\rangle}
{\langle\tilde{\epsilon}^2_{k_1,x}\rangle
\langle\tilde{\epsilon}^2_{k_2,x}\rangle}.
\end{equation} 
where the correlation is between wavelet coefficients at different scales but
at the same physical location. From Eqs. (21) and (22), we have approximately 
that
\begin{equation}
C(k_1,k_2,x) - 1 \sim
\left [
\frac{\langle \phi(t_{c1})\phi^*(t_{c2})\rangle}{\langle \phi^2 \rangle}
\right ]^2.
\end{equation}
Since $k_c \leq  H$, the scale-scale correlation given by Eq. (17) is
very significant. From Eq. (3), the time difference of forming perturbations 
on scales $k$ and $2k$ is $\Delta Ht = \log 2$.  Hence, 
$\langle \phi(t_{c1} +\log 2/H)\phi^*(t_{c1})\rangle/
\langle \phi^2\rangle $ is comparable to 1 according to Eq. (17). Therefore
$C(k,2k,x)-1$ can also be as large as about 1 implying that long-time tails 
can produce scale-scale correlations on the same order as the temperature 
fluctuations.

The correlation (17) is localized, i.e., the fluctuations
$\phi({\bf x},t)$ and $\phi({\bf x},t_0)$ are at the same position
${\bf x}$ with size about $1/H^3$. Therefore,
\begin{equation}
\langle \phi({\bf x}, t)\phi^{*}({\bf x'}, t_0) \rangle \simeq 0,
\end{equation}
if ${\bf x}$ and ${\bf x'}$ are different. As a consequence,
one can predict that the scale-scale correlations at different 
positions will be zero, or
\begin{equation}
C(k_1, k_2, x_1-x_2) \equiv
\frac{\langle \tilde{\epsilon}^2_{k_1, x_1}
   \tilde{\epsilon}^2_{k_2, x_2}\rangle}
{\langle\tilde{\epsilon}^2_{k_1, x_1}\rangle
\langle\tilde{\epsilon}^2_{k_2, x_2}\rangle} \sim 1,
\end{equation}
where $x_1-x_2 \neq 0$. We have tested this prediction for the
Ly-$\alpha$ forests. The results show a good agreement with
Eq. (25)\cite{PFG}.

It has generally been held that mode-mode coupling effects are negligible 
when perturbations are small and the linear approximation holds. The 
justification for this view can be seen from the decay of perturbations 
described by the linearized Boltzmann equation, which without considering 
diffusion effects, always decay exponentially. However this result is no
longer true for processes in which diffusion-like long wavelength modes
are important. In this case, even though the short wavelength modes 
decay exponentially [Eq. (9)], the long wavelength modes occur during the 
long-time tail which leads to scale-scale correlations [Eq. (17)].
Furthermore scale-scale correlations also arise in the conventional approach
to inflation due to the loss of quantum coherence via the mechanism of
coarse-grained histories \cite{nam}.
As such, scale-scale correlations are a generic feature of the 
inflation scenario with a slow-roll regime. The detected 
$C(11^{\circ}, 22^{\circ}) \sim 2$\cite{PLGF} implies that first evidence of 
inflation induced scale-scale correlations may have been found.
Finally, the detection of scale-scale correlations provides a way to reveal 
the time correlations of the fluctuations during inflation.


\begin{thebibliography}{99}
\bibitem{KT} e.g. E. Kolb \& M.S. Turner, {\it The Early Universe},
   Addison-Wesley, San Francisco, (1989), and references therein.
\bibitem{PLGF} J. Pando, P. Lipa, M. Greiner, \& L.Z. Fang,
   {\it Astrophys. J.} {\bf 496}, 9, (1998).
\bibitem{PVF} J. Pando, D. Valls--Gabaud \& L.Z. Fang,
   {\it Phys. Rev. Lett.}, {\bf 81}, 4568, (1998).
\bibitem{g95a} M. Greiner, P. Lipa \& P. Carruthers, {\it Phys. Rev.}
 {\bf E51}, 1948, (1995).
\bibitem{FT} L.Z. Fang \& R. Thews, {\it Wavelet in Physics}, World
Scientific, Singapore, (1998).
\bibitem{PF} J. Pando \& L.Z. Fang, {\it Phys. Rev. E}, {\bf E57},
 3593, (1998).
\bibitem{ehv} M.H. Ernst, E.H. Hauge \& J.M.J. van Leeuwen, 
 {\it Phys. Rev. Lett.} {\bf 25}, 1254, (1970).
\bibitem{KB} e.g. T.R. Kirkpatrick \& D. Belitz, {\it J. Stat. Phys.}
{\it 87}, 1307, (1997); N. Goldenfeld, {\it Lectures on Phase Transitions 
 and the Renormalization Group}, Addison Wesley, Reading MS, (1992);
 R. R\'esibois \& M. de Leener, {\it Classical Kinetic Theory of Fluids},
 John Wiley \& Son, New York, (1977).
\bibitem{star} A.A. Starobinsky, in {\em Fundamental Interactions}
 (MGPI Press, Moscow, 1984), p. 55; S. J. Rey, {\it Nucl. Phys}. {\bf B284}, 
 706, (1987); M. Sasaki, Y. Nambu and K. Nakao, {\it Nucl. Phys.} {\bf B308}, 
 868, (1988). 
\bibitem{LF} W.L. Lee \& L.Z. Fang, {\it Phys. Rev.}, {\bf D 59}, 083503 
(1999)
\bibitem{BF} A. Berera \& L.Z. Fang, {\it Phys. Rev. Lett.} 
 {\bf 74}, 1912, (1995); W.L. Lee \& L.Z. Fang, {\it Int. J. Mod. Phys.}, 
 {\bf D6}, 305, (1997).
\bibitem{nam} e.g. Y. Nambu, {Phys. Lett.} {B 276}, 11, (1992), and references
therein; also L.Z. Fang, W. Lee \& J. Pando, in preparation.
\bibitem{PFG} J. Pando, L.Z. Fang \& M. Greiner, in preparation.

\end{thebibliography}
\end{document}